\def\year{2021}\relax
  \documentclass[letterpaper]{article}
  \usepackage{aaai21} 
  \usepackage{times} 
  \usepackage{helvet} 
  \usepackage{courier} 
  \usepackage[hyphens]{url} 
  \usepackage{graphicx} 
  \usepackage{graphicx}  
  \usepackage{natbib}
  \usepackage{caption}
\usepackage{balance}
\usepackage{booktabs} 
\usepackage{soul}
\usepackage{subfigure}
\usepackage{pbox}
\usepackage{epsfig,endnotes}
\usepackage{epstopdf}
\usepackage{adjustbox}
\usepackage{xspace} 
\usepackage{xcolor}
\usepackage{url} 
\usepackage{enumitem}
\usepackage{chato-notes}[hide]
\usepackage{mathtools}
\usepackage{marginnote}
\usepackage[hang,flushmargin]{footmisc}

\usepackage{listings}
\usepackage{xcolor}

\colorlet{punct}{red!60!black}
\definecolor{background}{HTML}{EEEEEE}
\definecolor{delim}{RGB}{20,105,176}
\colorlet{numb}{magenta!60!black}

\lstdefinelanguage{json}{
    basicstyle=\ttfamily\footnotesize,
    stepnumber=1,
    numbersep=8pt,
    showstringspaces=false,
    breaklines=true,
    frame=lines,
    literate=
     *{0}{{{\color{numb}0}}}{1}
      {1}{{{\color{numb}1}}}{1}
      {2}{{{\color{numb}2}}}{1}
      {3}{{{\color{numb}3}}}{1}
      {4}{{{\color{numb}4}}}{1}
      {5}{{{\color{numb}5}}}{1}
      {6}{{{\color{numb}6}}}{1}
      {7}{{{\color{numb}7}}}{1}
      {8}{{{\color{numb}8}}}{1}
      {9}{{{\color{numb}9}}}{1}
      {:}{{{\color{punct}{:}}}}{1}
      {,}{{{\color{punct}{,}}}}{1}
      {\{}{{{\color{delim}{\{}}}}{1}
      {\}}{{{\color{delim}{\}}}}}{1}
      {[}{{{\color{delim}{[}}}}{1}
      {]}{{{\color{delim}{]}}}}{1},
}

\title{Tracking Knowledge Propagation Across Wikipedia Languages}

\author{
      Rodolfo Valentim\footnote{Part of this work was done during Rodolfo's internship at the Wikimedia Foundation.}\textsuperscript{\rm 1},
      Giovanni Comarela\textsuperscript{\rm 2},
      Souneil Park\textsuperscript{\rm 3},
      Diego Saez-Trumper\textsuperscript{\rm 4}\\}

\affiliations {
    \textsuperscript{\rm 1}Politecnico di Torino, \textsuperscript{\rm 2}Federal University of Espírito Santo,   \textsuperscript{\rm 3}Telefonica Research, 
    \textsuperscript{\rm 4}Wikimedia Foundation
    rodolfo.vieira@polito.it,
    gc@inf.ufes.br,
    souneil.park@telefonica.com,
    diego@wikimedia.org 
}

\newcommand{\gnote}[1]{{\tiny{\textcolor{red}{#1}}}}

\begin{document}

\maketitle



\begin{abstract}
In this paper, we present a dataset of \textit{inter-language knowledge propagation} in Wikipedia. Covering the entire 309 language editions and 33M articles, the dataset aims to track the full propagation history of Wikipedia concepts, and allow follow up research on building predictive models of them. For this purpose, we align all the Wikipedia articles in a language-agnostic manner according to the concept they cover, which results in 13M propagation instances. To the best of our knowledge, this dataset is the first to explore the full inter-language propagation at a large scale. Together with the dataset, a holistic overview of the propagation and key insights about the underlying structural factors are provided to aid future research. For example, we find that although long cascades are unusual, the propagation tends to continue further once it reaches more than four language editions. We 
also find that the size of language editions is associated with the speed of propagation. We believe the dataset not only contributes to the prior literature on Wikipedia growth but also enables new use cases such as edit recommendation for addressing knowledge gaps, detection of disinformation, and cultural relationship analysis. 
\end{abstract}






\section{Introduction}
\sloppy
Increasing global connectivity and digital tools foster knowledge propagation in all aspects; frequency, reach, and directions. Wikipedia is one of the key platform for knowledge propagation, especially between its language editions. The unique features of Wikipedia, such as wide language coverage, extensiveness of topics, and large-scale collaborative participation, provide an opportunity to track the knowledge propagation taking in place of the platform. The structural factors within Wikipedia, for example, cultural/linguistic similarity \cite{eom2015interactions, samoilenko2016linguistic}, distribution of multi-lingual editors \cite{hale2014multilinguals}, topics of common international interests \cite{kim2016understanding} also suggest the possibility to model and predict knowledge propagation. 

\begin{figure}[ht]
\centering
    \includegraphics[width=\linewidth]{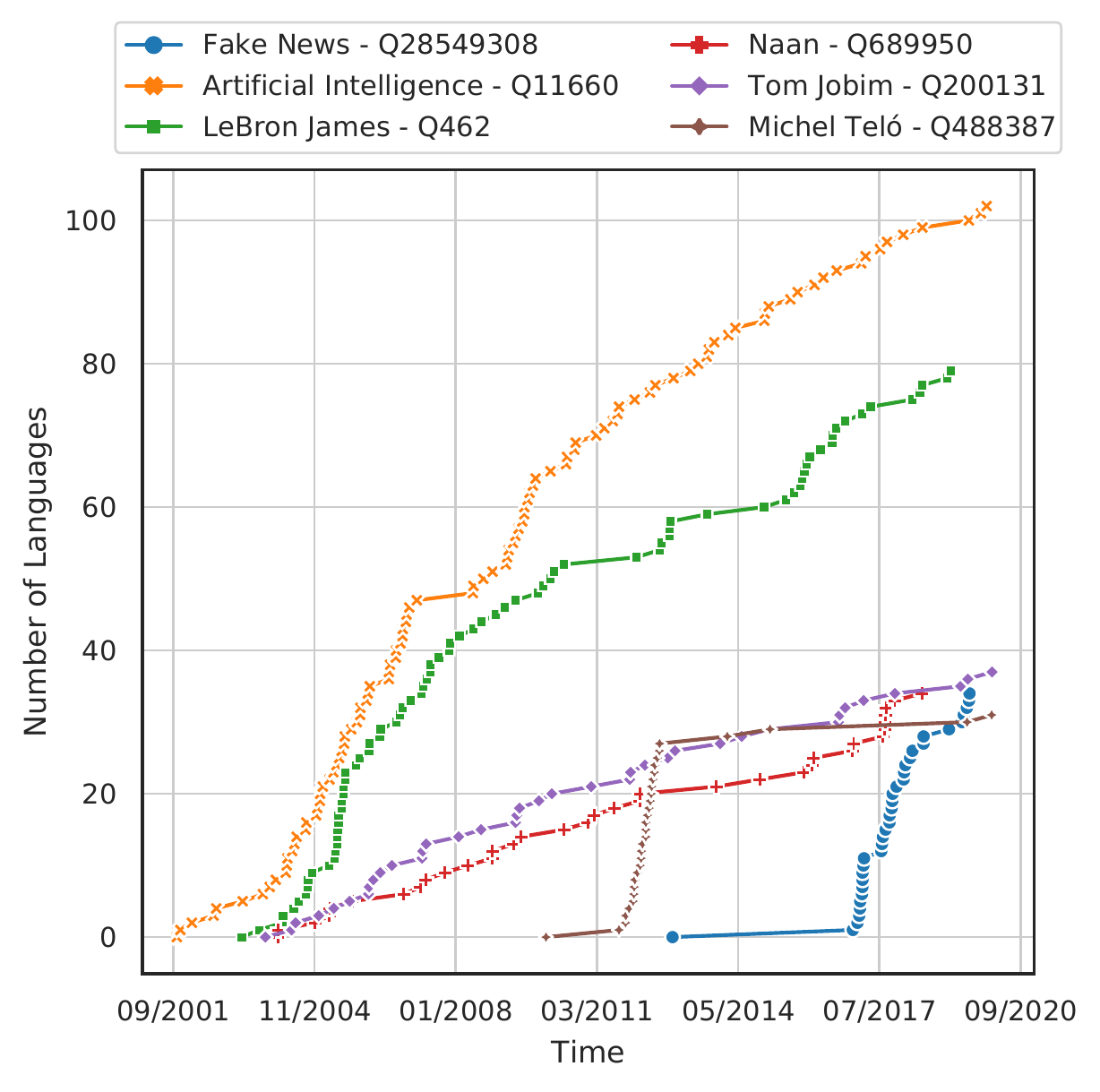}
    \caption{Example propagation across Wikipedia language editions.}
    \label{fig:example-cascades}
\end{figure}

In this paper, we develop a dataset to capture inter-language knowledge propagation in Wikipedia. 
The data set tracks knowledge propagation by aligning the entire page creations in a language agnostic manner. The sequence of page creations across the whole language editions is aggregated for every Wikipedia item. Figure~\ref{fig:example-cascades} shows some example propagation instances, a few showing a bursty expansion and a few showing a steady growth. Through this data set, we aim to provide a holistic view of the growth of Wikipedia from the perspective of inter-language propagation.



We believe that the data set has many application areas both within and beyond Wikipedia. First, it contributes to a deeper understanding of knowledge  propagation, which extends the literature on Wikipedia dynamics \cite{halfaker2013rise}, and supports data-driven modeling and prediction of knowledge propagation. This can lead to systems for expediting the growth Wikipedia, for example, that recommend pages for translation to editors based on the likelihood of propagation. Understanding the common propagation patterns could also help outlier detection, such as the spread of malicious content and cross-lingual sock-puppet attacks \cite{kumar2017army}. As Wikipedia languages reflect cultural boundaries and interactions, the data set can be used potentially in various cross cultural studies as well.

In order to promote the usages, we present various exploratory analyses together with the data set. 
For example, we observe that only a small fraction of items propagate to many language editions, consistent to the finding made in social networks \cite{goel2012structure}; however, propagation is more common in Wikipedia.
The propagation tends to gain speed, suggesting the possibility to predict whether an item would further propagate or not. We also find association between the size of language editions and the speed of propagation. We also observe more frequent propagation among language editions that show higher cultural and linguistic similarity, which corroborates the findings made in prior works on multi-lingual editors \cite{samoilenko2016linguistic}, coverage of popular figures \cite{eom2015interactions}.


The remainder of this paper is organized as follows: first, we briefly discuss the related work. We next   explain how we collected and developed the data set, and follow up with some basic statistics and characteristics of the data. Various uses cases are then elaborated, and the conclusion is presented with a number of future works. 



\section{Related Work}
\label{sec:related}
A large body of works was made on the evolution of Wikipedia. These include the work on measuring the rate of overall growth  \cite{almeida2007evolution}, discovery of the slowdown of the growth rate, and analyses of possible reasons \cite{suh2009singularity,halfaker2013rise}. Our work overlaps with this line of research as the data set captures the growth of Wikipedia; however, it provides a new perspective, i.e., inter-language propagation, to the growth. 

Study of the multiple language editions in Wikipedia provides related findings and suggests the possibility of growth through interactions between language editions. 
Hecht and Gergle focused on the 25 largest editions and observed that the majority of articles do not appear in many language editions: for example, 74\% of concepts are described in only one language edition \cite{hecht2010tower} (in our dataset, we find that number has decreased to 67\%). Hale's analysis \cite{hale2014multilinguals} looks into one of the key mechanisms of propagation, the edits made by multilingual editors. In the analysis, only a small set of users (15.4\%) is identified as multilingual editors. 
In fact, approaches to promote propagation across language editions have been made using various techniques including information retrieval, machine translation, and recommendation. Adar et al. developed a combination of machine learning techniques for improving infoboxes (tables with summary data of a page) of four large language editions \cite{adar2009information}. 
wikiBABEL employs machine translation and collaborative editing tools in order to facilitate content propagation across language editions \cite{kumaran2008wikibabel}. Wulczyn et al. presents a recommendation system for identifying missing articles and making personalized recommendations to editors \cite{wulczyn2016growing}. 
A deeper understanding of the inter-language propagation could aid such efforts and promote the interactions in a more efficient manner.


Similar types of studies on propagation have been actively conducted in diffusion research, including the epidemic of a disease \cite{anderson1992infectious,netrapalli2012learning}, diffusion of information in online social networks \cite{cheng2014can}, adoption of innovation \cite{leskovec2007dynamics} and the diffusion of scientific ideas \cite{cao-etal-2020-will}. Although Wikipedia does not have an explicit network structure and the propagation of our study is at the level of language editions rather than individuals, we share the view of diffusion research that the propagation can be modeled and predicted by learning the underlying patterns. 

\section{Dataset Development}
\label{sec:collection}

\subsection{Operationalization of knowledge propagation}

Content propagation across language editions could be studied through different types of edits, for example, text additions and revisions, image usages, or references, section restructuring, etc. 
The space of possible analysis is large as there are many types of edits. As a first step, our data set focuses on page creations. Page creations provides a practical means to view inter-language propagation since the alignment of the creations across language editions, \textit{i.e.}, identifying pages of the same concept, can be done clearly. Aligning other types of edits, such as text revisions, is more ambiguous and challenging as the content could diverge in many different ways through addition, deletion, paraphrasing, etc. In addition, the page creations are important since they are a seminal type of edits that opens a space for future updates. Thus, in our data set, we operationalize propagation as a sequence of page creations for the same Wikipedia concept among the language editions. 


Wikipedia is a huge corpus of densely inter-linked articles, not merely a simple collection of documents. 
The richness and diversity of its links allows the alignment of page creations in various ways.
In addition to the well-known links between articles (a.k.a \emph{pagelinks}), nowadays over 97\%  of Wikipedia articles are linked to Wikidata.\footnote{\url{https://wmdeanalytics.wmflabs.org/WD_percentUsageDashboard/}} Wikidata is a structured knowledge base where every concept is represented as a \emph{Wikidata item}, of which features include a collection of articles that describe the concept. 
Wikidata is language agnostic, an important feature which enables it to act as a centralized repository for all Wikipedia languages; inter-language links among equivalent articles can be extracted. 
For example, the article about \emph{Artificial Intelligence} in English Wikipedia, links to the Wikidata Item \emph{Q11660}, that is the same item that corresponds to \emph{Inteligencia Artificial} in Spanish Wikipedia, and to \emph{Kunstig intelligens} in the Danish version.  



By harnessing Wikidata items as language-agnostic concepts that appear in different Wikipedias, we can understand the interactions, similarity and differences among language editions. 
We use Wikidata items as an abstraction layer, and treat the article about an item as a language instance of it. 
Following the example above, we represent the article on \emph{Artificial Intelligence} in the English Wikipedia as the English instance of the Wikidata item \emph{Q11660}, and 
associate to it the timestamp of article creation.
Repeating the same process for the articles of all languages, we produce a sequence sorted by the creation time. 

Note that this sequence representation does not imply a linear chain of influence. In other words, there is no means to assure that the $t_{n+1}$-th element is created as a consequence of or influenced by the $t_n$-th element (or the ones before).
Here we only know the order, not causality. Apart from the small set of pages that are explicitly created through the translation tool,\footnote{\url{https://www.mediawiki.org/wiki/Content_translation}} 
we do not know the source nor the influence made to the creation of the pages. 

\subsection{Data extraction}

The dataset \cite{content_dataset2021}\footnote{The dataset can be downloaded directly from: \url{https://doi.org/10.5281/zenodo.4433137}} is developed by going through the following steps:

\begin{enumerate}
    \item Take all Wikipedia articles in all languages, and extract their date of creation. This is done through the public Wikipedia dumps.\footnote{\url{https://dumps.wikimedia.org}}
    \item Using the Wikidata dumps, map all articles to their corresponding Wikidata item.
    \item Next, group articles according to their Wikidata item, and create a sequence based on the the creation time. 
    \item For each sequence, we apply a topic model, classifying them over 64 hierarchical topic taxonomy. Note that an item can belong to more than one topic. 
    \item Finally, remove all the content created by bots. 
\end{enumerate}

\begin{figure*}[t]
    \centering
    \subfigure[Length of propagation.]{
	\includegraphics[scale=0.9]{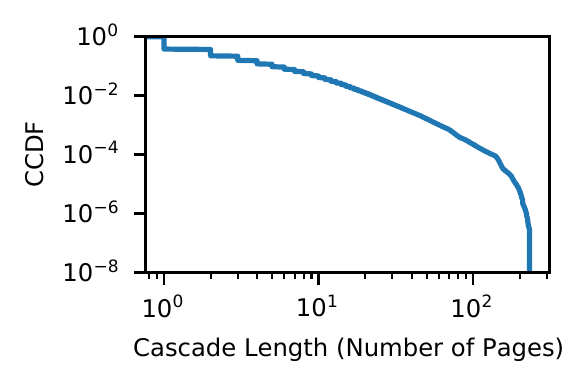}
	\label{fig:leng_dist}}
    \subfigure[Size of language editions.]{
	\includegraphics[scale=0.9]{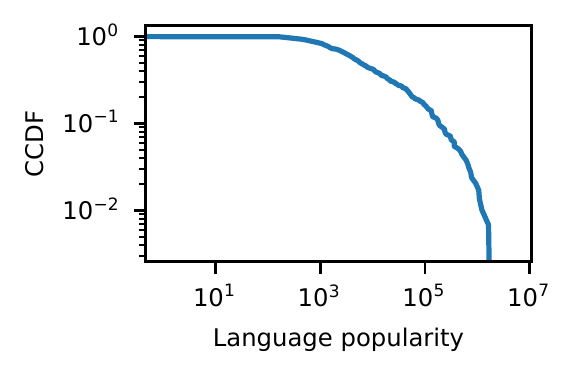}
	\label{fig:lang_size_dist}}
	\subfigure[Time interval between page creations.]{
	\includegraphics[scale=0.9]{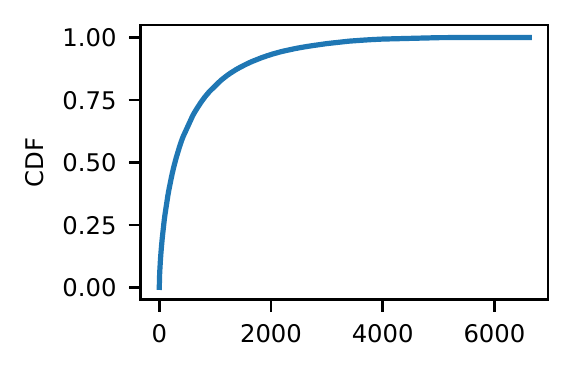}
	\label{fig:interval_dist}}
    \caption{Basic distributions of the data.}
    \label{fig:basicdesc}
\end{figure*}

\begin{lstlisting}[caption={JSON record representing Wikidata item with identifier Q2462783.},captionpos=b,label={lst:json},language=json,firstnumber=1]
{
    "wikidata_id": "Q2462783", 
    "editions": {
        "enwiki": 1070560507, 
        "nlwiki": 1330883168, 
        "fawiki": 1351061927, 
        "ptwiki": 1424133378, 
        "ruwiki": 1544017388, 
        "zh_yuewiki": 1562334127
    },
    "topics": {
        "STEM.STEM*": 0.96, 
        "STEM.Technology": 0.73
    }
}
\end{lstlisting}

The topics of each Wikidata item is identified through the Wiki-Topic tool.\footnote{https://wiki-topic.toolforge.org} The tool maps the items to a taxonomy of 64 hierarchical categories, including meta-topics like: Culture, Geography and STEM. 

We remove bots as we are interested in understanding the influence among editors of different languages, and also because there are few Wikipedias with high bot activity (such as the Cebuano Wikipedia) that introduce noise in our analysis. To remove bots, we use the public profile of the user who created each article. Note that in Wikipedia, \emph{bot users} are normally self declared; users who create accounts for bots (or uses bots within their own account) explicitly declares the information. 
While there are other techniques for bot detection we apply the standards defined by the Wikimedia community to delineate the edits made by bots.\footnote{\url{https://www.mediawiki.org/wiki/Manual:Bots}}



\subsection{Dataset format}

The dataset includes the data from 2001 to the first trimester of 2020. We first introduce some terminology.

\begin{itemize}
    \item \emph{Item}: the language-agnostic Wikidata identifier; for instance Q298 (Chile).\footnote{\url{https://www.wikidata.org/wiki/Q298}} 
    \item \emph{Page}: a specific language instance of an item. For example, the Portuguese version of Q298 would be ptwiki-Q298.\footnote{\url{https://pt.wikipedia.org/wiki/Chile}}
    \item \emph{Topic}: a set of items belonging to the same topic (e.g. History). Note that the topic is assigned to the item, and is propagated to all the pages of the item. Therefore, if item Q298 belongs to the topic 'Geography', all the pages about Q298 would also belong to the same topic.\footnote{\url{https://meta.wikimedia.org/wiki/Research:Language-Agnostic_Topic_Classification}}
\end{itemize}

Then, we organize our dataset as a collection of JSON-formatted records, one for each Wikidata item (See Listing \ref{lst:json} for an example). Following the definitions above, each record has three attributes:
\begin{itemize}
    \item \emph{wikidata\_id:} the Wikidata item;
    \item \emph{editions:} the list of pages, i.e., languages, related to an item and the creation time of each page.
For the page, the Wikimedia convention is to use the language code followed by the expression ``wiki'': for example, the English language is ``enwiki''.\footnote{The full list of the mapping between the code and the language can be found at \url{https://meta.wikimedia.org/wiki/Table_of_Wikimedia_projects}}
The creation time is represented as a 32-bit integer Unix timestamp.
    \item \emph{topics:} The list of topics that the content belongs to. Each topic is associated to a score ranging from 0 (not relevant) to 1 (relevant). We kept only the topics with score at least 0.5.
\end{itemize}

We emphasize that we removed non-Wikipedia projects such as Wiktionary, Wikiquote, Wikibooks, and others. Furthermore, we also make the same dataset available in a CSV file, which is sorted first by Wikidata item and then by time of page creation. We provide a Jupyter Notebook with examples on how to read and manipulate the dataset.\footnote{\url{https://github.com/rodolfovalentim/wikipedia-content-propagation}}


\section{Exploratory overview}
\label{sec:overview}

Figure \ref{fig:basicdesc} provides a basic characterization of propagation. In principle, we observe a skew of data in multiple aspects as expected: the length of propagation, size of language editions, and time interval between page creations.
We further look into possible patterns by exploring various potential associations between propagation, language editions, and topics. 

\textbf{Propagation length and momentum}: Figure \ref{fig:leng_dist} shows the distribution of propagation lengths, where most of the Wikidata items propagate to only a few language editions: for example, 88\% of the items propagate to less than five language editions. Although the distribution is skewed, we find that the skew is less than those observed in information diffusion made over social networks: for example, less than 1\% of the diffusion had greater diffusion tree depth than 2 \cite{goel2012structure}.

\begin{figure}[t]
\centering
    \includegraphics[width=\linewidth]{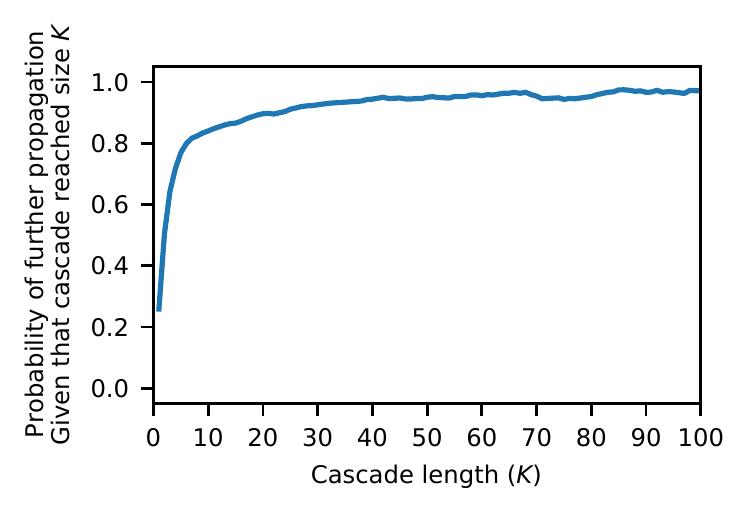}
    \caption{Probability of continue propagating given a cascade length $K$.}
    \label{fig:Probability-transition}
\end{figure}

We also observe that the propagation tends to gain momentum as it reaches more language editions. At every propagation length, we compute the ratio of the Wikidata items that reach another language edition to those who do not. Figure \ref{fig:Probability-transition} shows a rapid increase of the ratio, for example, from the third propagation, the ratio becomes greater than 0.5.

\begin{figure}[t]
\centering
    \includegraphics[width=\linewidth]{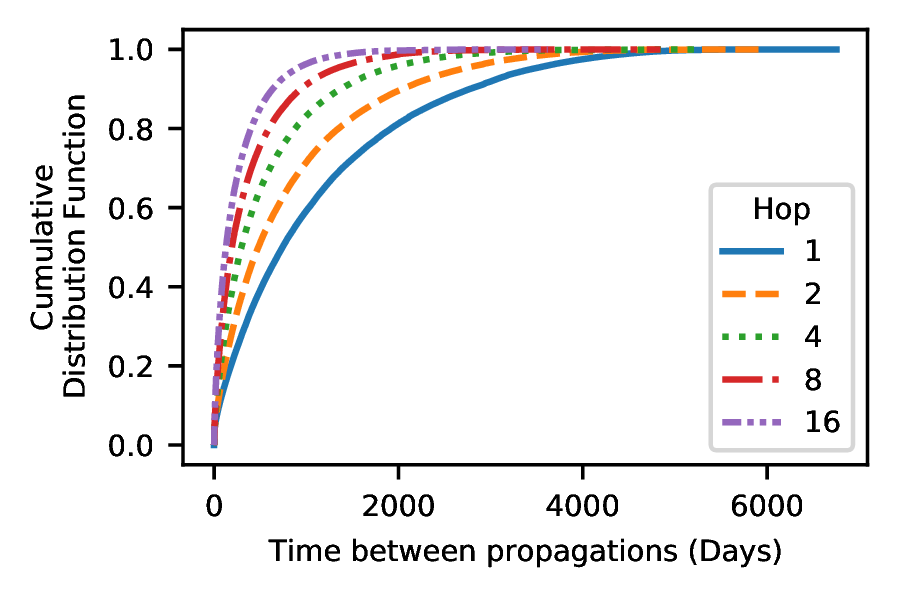}
\caption{Cumulative Distribution of the number of days between consecutive propagation relative to the position in the cascade. }
\vspace{-4mm}
\label{fig:intervals_by_hop}
\end{figure}

\textbf{Propagation speed and languages}: While we do not see a clear relationship between popular language editions and propagation length, we find associations between popular language editions and the speed of propagation. We first observe that the propagation tends to speed up as an item propagates further. The time interval between consecutive page creations for every item was measured and broken down by hop count. Figure \ref{fig:intervals_by_hop} depicts the distribution for a few example hops, showing that at greater hops, the time intervals are more skewed towards shorter values. This finding implies that an item is less likely to propagate further if a propagation event does not happen for a long time, and that this tendency gets stronger as the propagation progresses. 

\begin{figure}[t]
    \centering
 \includegraphics[width=\linewidth]{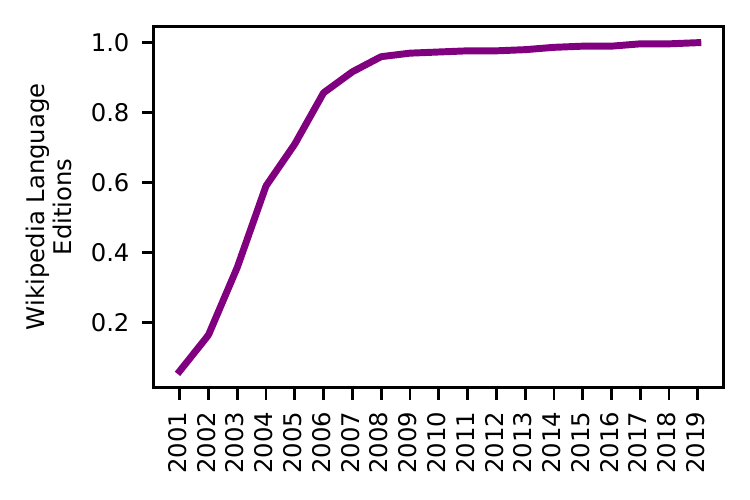}
    \caption{Number of Languages covered by Wikipedia.}
    \label{fig:langsPerYear}
\end{figure}

\begin{figure*}[ht]
\includegraphics[width=\linewidth]{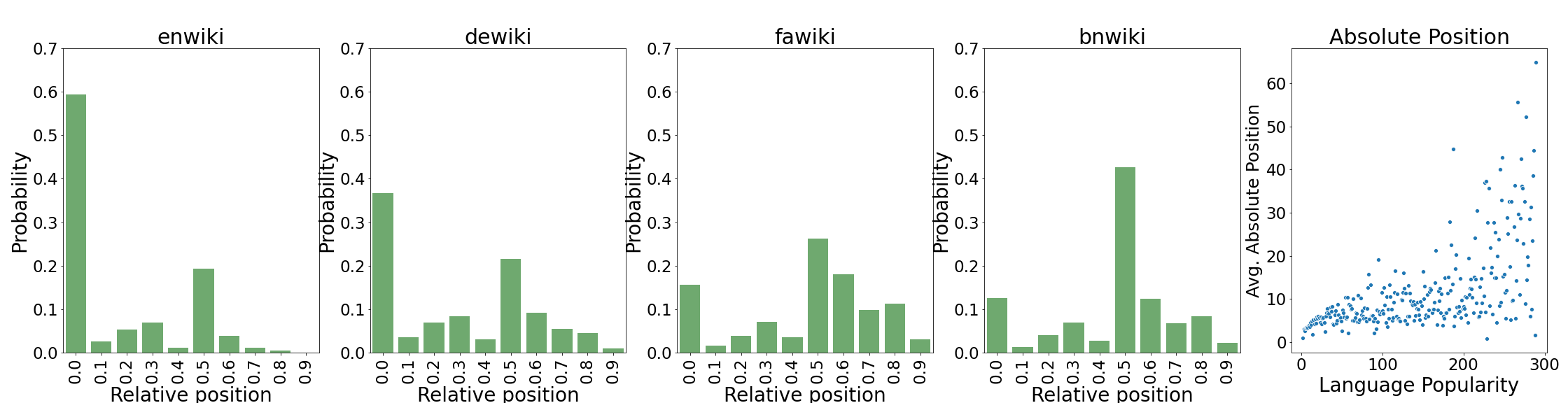}
\caption{Distribution of relative order in propagation\label{fig:rel_order_dist}.}
\end{figure*}

In addition to this finding, we find that popular language editions tend to appear early in the propagation sequence. To avoid the bias coming from the longer history of popular language editions, we only consider the propagation that started after year 2008, when most of the language editions were present in Wikipedia (Figure~\ref{fig:langsPerYear} shows the evolution of Wikipedia in terms of language coverage). For each language edition, we compute and aggregate the relative position in the propagation (i.e., its order divided by the length of propagation). The histograms in Figure \ref{fig:rel_order_dist} shows the distribution of the relative position for two example large language editions (i.e., English and German) and two smaller language editions (i.e., Farsi, Bengali). It shows that the distribution of large language editions is more skewed to smaller values. The scatter plot of Figure \ref{fig:rel_order_dist} provides an overview of all language editions, sorting the language editions along the x-axis based on their size and plotting their average absolute position in the propagation. The results suggest an hypothesis for further examination and exploration of causal relationships, i.e., if an item being covered in popular language editions reduces the time of creations in other language editions. 

\begin{figure}[t]
\centering
    \includegraphics[width=0.9\linewidth]{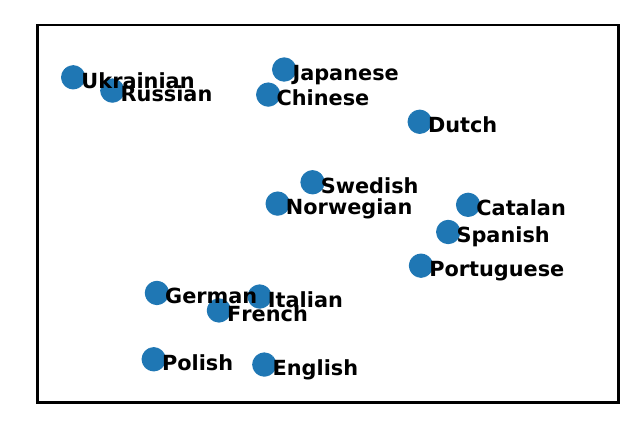}
\caption{Ties among example language editions in propagation 
}
\label{fig:jac_idx}
\end{figure}

\textbf{Inter-language ties}: We also observe ties between language editions in the propagation process, which reflect linguistic and cultural relationships among them. For every pair of language editions, we compute the Jaccard index\footnote{Jaccard index for two sets, A and B, is computed as $J(A,B) = |A\cap B| / |A \cup B|$} of the set of items they cover. Figure \ref{fig:jac_idx} provides an overview of the distance among language editions. Based on the Jaccard index values, the languages are projected to a 2D space using t-SNE~\cite{tsne}. 

We additionally compare our inter-language edition analysis with the translation activities of editors assuming that page translations are an important means of propagation in Wikipedia. Taking the logs of the Wikipedia translation tool\footnote{\url{https://www.mediawiki.org/wiki/Content_translation}}, we measure the correlation between the computed Jaccard index and the frequency of translation activities between the language editions. A moderately high correlation (Spearman's Rho $r_s = 0.41$ , $p < .01$) suggests that direct page translation is one of the key mechanism of propagation, however, at the same time, the number suggests that a significant part is not directly associated to translation. Furthermore, the translation activities did not exist for a vast majority of language pairs (98\%), hence those pairs had to be excluded from the correlation analysis.

\section{Use Cases}

In general, this data set provides an integrated historical view of Wikidata and Wikipedia, making ease of access and creating synergy of the two sources. 

There are multiple areas of applications. First of all, 
the data set can directly aid researchers conducting multilingual studies in Wikipedia.
While there has been a wide set of research focusing on English Wikipedia, relatively less has been done on the interaction between different languages. Particularly, our data set helps measuring temporal page co-occurrence patterns among language editions, which can expand the studies on distances and similarities of Wikipedia editions \cite{eom2015interactions,kim2016understanding}. The data set allows going into more details by segmenting the data by topic or by range of propagation. Furthermore, the time span covered ($\sim$20 years) and the time granularity of each data point enables longitudinal studies or focusing on specific periods of time.

The possibilities for developing machine learning models is also wide. For instance, it would be interesting to approach the problem of predicting the propagation sequence, \textit{i.e.,}, to which new language a given item will propagate or the final cascade length. The models can provide insights on  frequent propagation patterns and key features of them. 

The models can lead to important use cases, for example, addressing knowledge gaps among the language editions. Recommender systems that suggest articles for translation or topics to editors of a target language based on the propagation status can be built with the models. On the other hand, the models may also assist detecting disinformation that spread to many language editions in an unusual manner. 

It is important to highlight that the Wikidata identifiers\footnote{\url{https://www.wikidata.org/wiki/Help:Items}} further open the opportunity to enrich this data set with semantic information and allow more in-depth studies. For example, a researcher could dive into a particular set of items (e.g., items of large propagation) by collecting detailed information through Wikidata APIs (we give more details about the possibilities of connecting this data set with other databases in the next section). 

There are also applications beyond Wikipedia. 
Assuming language editions as a proxy of cultural boundaries, the data set can support various cross-cultural studies such as understanding cultural similarity, interactions, and their evolution over time.
Comparisons against social media and Wikipedia could be also made to understand the differences between knowledge propagation and information diffusion.

\section{Ethical \& FAIR Considerations}

This work is entirely based on public data, and does not contain personal information. 

This data set is developed in accordance with the \emph{FAIR Data Principles}. The objects and attributes of the data set are clearly defined. As mentioned in the previous section, it can also be enriched with various metadata available in  Wikidata. The extensive set of external identifiers of Wikidata (e.g., authority controls, international standard identifiers, and social media ids\footnote{\url{https://www.wikidata.org/wiki/Wikidata:List_of_properties/Wikidata_property_for_an_identifier}}) also makes the data set interoperable and re-usable. Researchers can easily combine the data set with other databases through the identifiers. 

The nature of Wikidata as a free knowledge base of linked data allows re-using our work in different scenarios, applications, and as core component or complementary data for future studies.


\section{Conclusions}\label{sec:conc}

In this paper, we present a dataset of inter-language knowledge propagation in Wikipedia. 
The dataset captures the propagation history of Wikipedia by aligning the whole pages of all language editions and sorting them by the page creation time. Users can conduct various analyses, for example, temporal patterns in page creations between language editions, impact of particular language editions or topics, and different evolution processes regarding the speed and range of propagation. We provide a statistical overview and a number of potential patterns in order to guide follow up research. 


There are diverse applications of the dataset. Methods for efficiently transferring knowledge from one Wikipedia community to another could be explored using the dataset, and mitigate the problem of knowledge gaps. Machine learning models for detecting hot content in Wikipedia with cross-cultural interest can be created. Similar models can  be developed to detect unusual content propagation, potentially driven by malicious users or coalitions trying to introduce a specific topic or point of view in Wikipedia. 
Given the large amount of languages covered in the data set and the relevance of Wikipedia in the global knowledge ecosystem, the data can be also used outside of Wikipedia, for example, to understand cultural similarities or enhance multilingual NLP systems.

There are multiple directions for future work. A direct extension is to develop a prediction model of the propagation. We also plan to enrich the data set, for example, by incorporating page views, and advance the prediction model accordingly. Including page views adds another dimension and allows new analyses on the interplay between content popularity and propagation. As mentioned early on in the paper, propagation could be also studied with different types of edits and at finer grained levels than pages, for example, images and sections. 


\section{Acknowledgments}
We acknowledge the Research Internship program at Wikimedia Foundation for founding Rodolfo Valentim's internship. Giovanni Comarela is financed in part by CAPES (Finance  Code  001), CNPq, and FAPES.

\bibliography{KP.bib}


\end{document}